\documentstyle[pra,aps,epsf]{revtex}

\begin{document}
\thispagestyle{empty}

\renewcommand {\thefootnote}{\fnsymbol{footnote}}
\renewcommand{\thesection}{\Roman{section}}
\def \beq {\begin{equation}}
\def \eeq {\end{equation}}
\def \bes {\begin{eqnarray}}
\def \ees {\end{eqnarray}}
\def\rv{\mbox{\boldmath$r$}}
\def\drv{{d\mbox{\boldmath$r$}}}
\def\mum {$\,\mu\mbox{m}$}
\draft
\thispagestyle{empty}
\title
{ Casimir and van der Waals force between two plates
or a sphere (lens) above a plate made of real metals}

\author
{
G.~L.~Klimchitskaya,$\!{}^{1,}$\footnote{On leave from 
North-West Polytechnical Institute,
St.Petersburg, Russia.
Electronic address: galina@GK1372.spb.edu}${}^{\S}$
U.~Mohideen,$\!{}^{2,}$\footnote{Electronic address:
umar.mohideen@ucr.edu}
and V.~M.~Mostepanenko$\!{}^{1,}$\footnote{On 
leave from
A.Friedmann Laboratory for Theoretical Physics,
St.Petersburg, Russia. Electronic address:
mostep@fisica.ufpb.br}\footnote{Present Address: Institute for Theoretical 
Physics,
Leipzig University, Augustusplatz 10/11, 04109, Leipzig, Germany.}}
\address
{
${}^1\,${ Physics Department, Federal University
of Para\'{\i}ba, C.P.5008, \\
CEP 58059--970, Jo\~{a}o Pessoa, Pb---Brazil}
\\[3mm]
${}^2\,${ Department of Physics, University of
California, Riverside,
\\California 92521}
}

\maketitle

\begin{abstract}
The Casimir and van der Waals forces acting between two 
metallic plates or a sphere (lens) above a plate are calculated 
accounting for the finite conductivity of the metals. The simple 
formalism of surface modes is briefly presented which allows the 
possibility to obtain the generalization of
Lifshitz results for the case of two semi-spaces covered by the thin layers.
Additional clarifications of the regularization procedure provides the
means to obtain reliable results not only for the force but also for the
energy density. This, in turn, leads to the value of the force
for the configuration of a sphere (lens) above a plate 
both of which are covered by
additional layers. The Casimir interaction between $Al$ and $Au$ test
bodies is recalculated using the optical tabulated data for the
complex refractive index of these metals. The computations turn out to be
in agreement with the perturbation theory up to the fourth order in
relative penetration depth of electromagnetic zero point  os\-cil\-la\-tions 
into the
metal. The disagreements between the results recently presented 
in the literature
are resolved. The Casimir force between 
$Al$ bodies covered by the thin
$Au$ layers is computed and the possibility to neglect spatial
dispersion effects is discussed as a function of the layer thickness. The 
van der 
Waals force is calculated including the transition region to the Casimir
force. The pure non-retarded van der Waals force law between $Al$ and $Au$
bodies is shown to be restricted to a very narrow distance interval
from 0.5\,nm to (2--4)\,nm. New, more exact, values of the Hamaker
constant for $Al$ and $Au$ are determined.
\end{abstract}

\vskip 8mm

\section{INTRODUCTION}

Recently considerable attention has been focussed on the van der Waals
and Casimir forces acting between macroscopic bodies. As for the van der
Waals force, interest in it has quickened owing to its application in
atomic force microscopy (see, e.g., the monographs \cite{1,2} and
references therein). Interest in the Casimir force was rekindled after
the new experiments \cite{3,4} where it was measured more precisely
in the case of metallic test bodies.

It is common knowledge that both forces are connected with the
existence of zero point vacuum oscillations of the electromagnetic 
field \cite{5,6}.
For closely spaced macroscopic bodies the virtual photon emitted by an atom
of one body reaches an atom of the second body during its lifetime. The
correlated oscillations of the instantaneous induced dipole moments of those
atoms give rise to the non-retarded van der Waals force. The Casimir force
arises when the distance between two bodies is so large that the virtual
photon emitted by an atom of one body cannot reach the second body during
its lifetime. Nevertheless, the correlation of the quantized electromagnetic
field in a vacuum state is not equal to zero at two points where the atoms
belonging to different bodies are situated. Hence the non-zero correlations 
of the induced atomic dipole moments arise once more resulting
in the Casimir force (which is also known as the retarded van der Waals
force).

As is shown in \cite{4,7,8}, the corrections to the Casimir force due to
the finite conductivity of the metal and surface roughness play an
important role in the proper interpretation of the measurement data.
Temperature corrections are negligible in the measurement range of
\cite{4,7,8} (data of \cite{3} do not support the presence of finite
conductivity, surface roughness and temperature corrections which is in 
disagreement with the theoretically estimated values of these corrections
\cite{7} in the measurement range of \cite{3}). In \cite{4,7} the values
of the finite conductivity corrections to the Casimir force were found by
the use of perturbation expansion in relative penetration depth of
electromagnetic zero point oscillations into the metal which starts from 
the general
Lifshitz formula [9--11]. The parameter of this expansion is
$\lambda_p/(2\pi a)$, where $\lambda_p$ is the effective plasma frequency
of the electrons, $a$ is the distance between interacting bodies. Note
that the coefficient near the first order correction was obtained in
\cite{12,13}, and near the second order one in \cite{14} for the configuration
of two plane parallel plates. In \cite{3,15} the results of \cite{12,13}
and, correspondingly, \cite{14} were modified for the configuration of a
spherical lens above a plate. To do this the proximity force theorem
\cite{16} was applied. The coefficients to the third and
fourth order terms of that expansion were first obtained in \cite{17} for both
configurations.

In applications to atomic force microscopy and the van der Waals force the
Lifshitz formula and plasma model were used in \cite{18,19} for different 
configurations of a tip above a plate. In \cite{20,21}, the 
density-functional theory along with the
plasma model was used in the calculation of the van der Waals force. 
More complicated analytical representation for the
dielectric permittivity (Drude model with approximate account of absorption
 bands) 
was used in \cite{22} to calculate the van der Waals force between objects 
covered with a chromium layer with the Lifshitz formula.

The parameters of plasma and Drude models (plasma wavelength, electronic
relaxation frequency) are not known very precisely. Due to this in
\cite{23} the attempt was undertaken to apply Lifshitz formalism
numerically to gold, copper, and aluminum (see also \cite{24}).
The tabulated data for the frequency dependent complex refractive
index of these metals were used together with the dispersion
relation to calculate the values of dielectric permittivity on the
imaginary frequency axis. Thereupon the Casimir force was calculated in
\cite{23} for configurations of two plates and a spherical lens above a plate
in a distance range from 0.05{\mum} to 2.5{\mum}. The same computation
based on Lifshitz formalism and optical tabulated data for the dielectric
permittivity was repeated in \cite{25} in a distance range from 0.1{\mum}
to 10{\mum}. The two sets of results are in
disagreement (see also \cite{26}). 
Note that the higher-order perturbative
calculations of \cite{17} in their application range are in
agreement with \cite{25,26} but also disagree with \cite{23,24}.

In this paper we present a brief derivation of the van der Waals and
Casimir energy density and force between two parallel
metallic plates or a plate
and a sphere covered by the thin layers of another metal (the configuration
used in the experiments \cite{4,8}). Two plates of sufficient thickness
can be modelled by two semi-spaces with some gap between them.
The case of multilayered plane walls
was considered in \cite{27}. In contrast to \cite{27} where the removal of
the infinities of the zero-point energy was not considered, we present
explicitly the details of the regularization procedure and its physical
justification. We next perform an independent computation using optical 
tabulated data for the frequency dependent complex 
refractive index of
aluminum and gold with the goal to resolve the disagreement between 
earlier results.
Our results turn out to be in agreement with \cite{25,26} with a precision
of computational error less than 1\%. Also the influence of the thin covering
metallic layers onto the Casimir force is determined. The range of 
applicability and exceptions to using
the bulk metal optical data for the dielectric permittivity of the thin 
metallic layers is discussed. For smaller distances 
the intermediate (transition) 
region between the Casimir and van der Waals forces is examined. It is shown
that the transition region  is very wide ranging from
several nanometers to hundreds of nanometers. The pure van der Waals regime
for aluminum and gold is restricted to separations in the interval from 0.5\,nm
till (2--4)\,nm only. The more exact values of the Hamaker constant for
aluminum and gold are determined with the use of obtained computational data.

The paper is organized as follows. In Sec.~II the general formalism is briefly
presented giving the Casimir and van der Waals forces including the effect 
of covering layers on the surface of interacting bodies (two plates or a 
sphere above a plate). In Sec.~III the influence of finite conductivity 
of the metal onto the Casimir force is reexamined.
Sec.~IV contains the calculation of the
Casimir force between the aluminum surfaces covered by the thin gold layers.
In Sec.~V the van der Waals force is calculated in both configurations 
and the transition region to the Casimir is examined.
Sec.~VI contains determination of the Hamaker constant values for aluminum
and gold. In Sec.~VII we present conclusions and discussion, in
particular, of possible applications of the obtained results 
in experimental investigations of the Casimir force and for obtaining stronger
constraints on the constants of hypothetical long-range interactions. 

\section{THE VAN DER WAALS AND CASIMIR FORCE BETWEEN LAYERED SURFACES:
GENERAL FORMALISM}

We consider first two semi-spaces bounded by planes ($x,y$) and filled
with material having a frequency-dependent dielectric permittivity
$\varepsilon_2(\omega)$. Let the planes bounding the semi-spaces be covered
by layers of thickness $d$ made of the another material with a dielectric
permittivity $\varepsilon_1(\omega)$. The magnetic permeabilities of both
materials are taken to be equal to unity. The region of thickness $a$
between the layers
 (see Fig.~1) is empty space. According to \cite{28,29}
van der Waals and Casimir forces for the configuration under consideration can
be found by consideration of the surface modes for which
$\mbox{div\boldmath$E$}=0$, 
$\mbox{curl\boldmath$E$}=0$. 
The infinite zero-point energy of electromagnetic field, dependent on $a$
and $d$,
is given by \cite{5,27}
\beq
E(a,d)=\frac{1}{2}\hbar
\sum_{\mbox{\boldmath$\scriptstyle{k}$},n}
\left(\omega_{\mbox{\boldmath$\scriptstyle{k}$},n}^{(1)}+
\omega_{\mbox{\boldmath$\scriptstyle{k}$},n}^{(2)}\right).
\label{1}
\eeq
\noindent
Here $\omega_{\mbox{\boldmath$\scriptstyle{k}$},n}^{(1,2)}$ are the proper
frequencies of the surface modes with two different polarizations of the
electric field (parallel and perpendicular to the plane formed by
{\boldmath$k$} and $z$ axis correspondingly), {\boldmath$k$} is the
two-dimensional propagation vector in the $xy$-plane.

For the vacuum energy density per unit area of the bounding planes (which is
also infinite) one obtains from (\ref{1})
\beq
{\cal E}(a,d)=
\frac{E(a,d)}{L^2}=\frac{\hbar}{4\pi}
\int\limits_{0}^{\infty}k\,dk\,
\sum_{n}
\left(\omega_{\mbox{\boldmath$\scriptstyle{k}$},n}^{(1)}+
\omega_{\mbox{\boldmath$\scriptstyle{k}$},n}^{(2)}\right),
\label{2}
\eeq
\noindent
where $L$ is the side-length of bounding plane.

The frequencies of the surface modes
 $\omega_{\mbox{\boldmath$\scriptstyle{k}$},n}^{(1,2)}$ 
are found from the boundary conditions for the electric field and magnetic
induction imposed at the points $z=-\frac{a}{2}-d$,  $-\frac{a}{2}$,
 $\frac{a}{2}$, and  $\frac{a}{2}+d$ \cite{27}.
These boundary conditions for each polari\-za\-tion lead to a system of
eight linear homogeneous equations. The requirements that these equations
have non-trivial solutions are
\bes
&&
\Delta^{(1)}\left(\omega_{\mbox{\boldmath$\scriptstyle{k}$},n}^{(1)}\right)
\equiv
e^{-R_2(a+2d)}\left\{\left(r_{10}^{+}r_{12}^{+}e^{R_1d}-
r_{10}^{-}r_{12}^{-}e^{-R_1d}\right)^2e^{R_0a}
-\left(r_{10}^{-}r_{12}^{+}e^{R_1d}-
r_{10}^{+}r_{12}^{-}e^{-R_1d}\right)^2e^{-R_0a}\right\}=0,
\label{3}\\
&&
\Delta^{(2)}\left(\omega_{\mbox{\boldmath$\scriptstyle{k}$},n}^{(2)}\right)
\equiv
e^{-R_2(a+2d)}\left\{\left(q_{10}^{+}q_{12}^{+}e^{R_1d}-
q_{10}^{-}q_{12}^{-}e^{-R_1d}\right)^2e^{R_0a}
-\left(q_{10}^{-}q_{12}^{+}e^{R_1d}-
q_{10}^{+}q_{12}^{-}e^{-R_1d}\right)^2e^{-R_0a}\right\}=0.
\nonumber
\ees
\noindent
Here the following notations are introduced
\beq
r_{\alpha\beta}^{\pm}=R_{\alpha}\varepsilon_{\beta}\pm
R_{\beta}\varepsilon_{\alpha}, \qquad
q_{\alpha\beta}^{\pm}=R_{\alpha}\pm R_{\beta},
\qquad
R_{\alpha}^2=k^2-\varepsilon_{\alpha}\frac{\omega^2}{c^2},
\qquad \varepsilon_0=1, \qquad \alpha=0,\,1,\,2.
\label{4}
\eeq
\noindent
Note that to obtain Eqs.~(\ref{3}) we set the determinants of the linear
system of equations equal to zero and do not perform any additional 
transformations.
This is the reason why (\ref{3}) does not coincide with the corresponding
equations of \cite{5,27} where some transformations were used which are
not equivalent in the limit $|\omega|\to\infty$ (see below).

Summation in (\ref{2}) over the solutions of (\ref{3}) can be performed
with the help of the argument 
principle which was applied for this purpose in
\cite{28}. According to this principle
\beq
\sum_{n}
\omega_{\mbox{\boldmath$\scriptstyle{k}$},n}^{(1,2)} =
\frac{1}{2\pi i}\left[
\int\limits_{i\infty}^{-i\infty}
\omega d\ln\Delta^{(1,2)}(\omega)
+
\int\limits_{C_{+}}
\omega d\ln\Delta^{(1,2)}(\omega)\right]
\label{5},
\eeq
\noindent
where $C_{+}$ is a semicircle of infinite radius in the right one-half of the
complex $\omega$-plane with a center at the origin. Notice that the
functions $\Delta^{(1,2)}(\omega)$, defined in (\ref{3}), have no poles.
For this reason the sum over their poles is absent from (\ref{5}).

The second integral in the right-hand side of (\ref{5}) is simply
calculated with the natural supposition that
\beq
\lim\limits_{\omega\to\infty}\varepsilon_{\alpha}(\omega)=1,
\qquad
\lim\limits_{\omega\to\infty}
\frac{d\varepsilon_{\alpha}(\omega)}{d\omega}=0
\label{6}
\eeq
\noindent
along any radial direction in complex $\omega$-plane. The result is
infinite, and does not depend on $a$:
\beq
\int\limits_{C_{+}}
\omega\, d\ln\Delta^{(1,2)}(\omega)
=4\int\limits_{C_{+}}
d\omega. 
\label{7}
\eeq

Now we introduce a new variable $\xi=-i\omega$ in (\ref{5}), (\ref{7}).
The result is
\beq
\sum_{n}
\omega_{\mbox{\boldmath$\scriptstyle{k}$},n}^{(1,2)} =
\frac{1}{2\pi }
\int\limits_{\infty}^{-\infty}
\xi\, d\ln\Delta^{(1,2)}(i\xi)+
\frac{2}{\pi}
\int\limits_{C_{+}}
d\xi,
\label{8}
\eeq
\noindent
where both contributions in the right-hand side diverge. To remove the
divergences we use the regularization procedure which goes back to the
original Casimir paper \cite{30} (see also \cite{6,28}). The idea of this
procedure is that the regularized physical vacuum energy density vanishes
for the infinitely separated interacting bodies. From Eqs.~(\ref{3}), (\ref{8})
it follows
\beq
\lim\limits_{a\to\infty}
\sum_{n}
\omega_{\mbox{\boldmath$\scriptstyle{k}$},n}^{(1,2)} =
\frac{1}{2\pi }
\int\limits_{\infty}^{-\infty}
\xi\, d\ln\Delta_{\infty}^{(1,2)}(i\xi)+
\frac{2}{\pi}
\int\limits_{C_{+}}
d\xi,
\label{9}
\eeq
\noindent
where the asymptotic behavior of $\Delta^{(1,2)}$ at $a\to\infty$
is given by
\beq
\Delta_{\infty}^{(1)}=
e^{(R_0-R_2)a-2R_2d}
\left(r_{10}^{+}r_{12}^{+}e^{R_1d}-
r_{10}^{-}r_{12}^{-}e^{-R_1d}\right)^2,
\qquad
\Delta_{\infty}^{(2)}=
e^{(R_0-R_2)a-2R_2d}
\left(q_{10}^{+}q_{12}^{+}e^{R_1d}-
q_{10}^{-}q_{12}^{-}e^{-R_1d}\right)^2.
\label{10}
\eeq

Now the regularized physical quantities are found with the help of
(\ref{8})--(\ref{10})
\beq
\left(\sum_{n}
\omega_{\mbox{\boldmath$\scriptstyle{k}$},n}^{(1,2)}\right)_{reg}
\equiv
\sum_{n}
\omega_{\mbox{\boldmath$\scriptstyle{k}$},n}^{(1,2)}-
\lim\limits_{a\to\infty}
\sum_{n}
\omega_{\mbox{\boldmath$\scriptstyle{k}$},n}^{(1,2)}=
\frac{1}{2\pi }
\int\limits_{\infty}^{-\infty}
\xi\, d\ln\frac{\Delta^{(1,2)}(i\xi)}{\Delta_{\infty}^{(1,2)}(i\xi)}.
\label{11}
\eeq
\noindent
They can be transformed to a more convenient form with the help of
integration by parts
\beq
\left(\sum_{n}
\omega_{\mbox{\boldmath$\scriptstyle{k}$},n}^{(1,2)}\right)_{reg}=
\frac{1}{2\pi }
\int\limits_{-\infty}^{\infty}
d\xi\, \ln\frac{\Delta^{(1,2)}(i\xi)}{\Delta_{\infty}^{(1,2)}(i\xi)},
\label{12}
\eeq
\noindent
where the term outside the integral vanishes.

To obtain the physical, regularized Casimir energy density one should
substitute the regularized quantities (\ref{12}) into (\ref{2}) instead of
(\ref{8}) with the result
\beq
{\cal E}_{reg}(a,d)=\frac{\hbar}{4\pi^2}
\int\limits_{0}^{\infty} k\,dk
\int\limits_{0}^{\infty} d\xi
\left[\ln Q_1(i\xi)+ \ln Q_2(i\xi)\right],
\label{13}
\eeq
\noindent
where
\bes
&&
Q_1(i\xi)\equiv\frac{\Delta^{(1)}(i\xi)}{\Delta_{\infty}^{(1)}(i\xi)}
=1-\left(\frac{r_{10}^{-}r_{12}^{+}e^{R_1d}-
r_{10}^{+}r_{12}^{-}e^{-R_1d}}{r_{10}^{+}r_{12}^{+}e^{R_1d}-
r_{10}^{-}r_{12}^{-}e^{-R_1d}}\right)^2e^{-2R_0a},
\nonumber\\
&&
Q_2(i\xi)\equiv\frac{\Delta^{(2)}(i\xi)}{\Delta_{\infty}^{(2)}(i\xi)}
=
1-\left(\frac{q_{10}^{-}q_{12}^{+}e^{R_1d}-
q_{10}^{+}q_{12}^{-}e^{-R_1d}}{q_{10}^{+}q_{12}^{+}e^{R_1d}-
q_{10}^{-}q_{12}^{-}e^{-R_1d}}\right)^2e^{-2R_0a}.
\label{14}
\ees
\noindent
In (\ref{13}) $Q_{1,2}$ are even functions
of $\xi$ has been taken into account.

For the convenience of numerical calculations below we introduce the new
variable $p$ instead of $k$ defined by
\beq
k^2=\frac{\xi^2}{c^2}(p^2-1).
\label{15}
\eeq
\noindent
In terms of $p,\,\xi$ the Casimir energy density (\ref{13}) takes the form
\beq
{\cal E}_{reg}(a,d)=\frac{\hbar}{4\pi^2c^2}
\int\limits_{1}^{\infty} p\,dp
\int\limits_{0}^{\infty} \xi^2\,d\xi
\left[\ln Q_1(i\xi)+ \ln Q_2(i\xi)\right],
\label{16}
\eeq
\noindent
where a more detailed representation for the functions $Q_{1,2}$ from
(\ref{14}) is
{\normalsize
\bes
&&
Q_1(i\xi)=1-\left[\frac{(K_1-\varepsilon_1p)(\varepsilon_2K_1+
\varepsilon_1K_2)-(K_1+\varepsilon_1p)(\varepsilon_2K_1-\varepsilon_1K_2)
e^{-2\frac{\xi}{c}K_1d}}{(K_1+\varepsilon_1p)(\varepsilon_2K_1+
\varepsilon_1K_2)-(K_1-\varepsilon_1p)(\varepsilon_2K_1-\varepsilon_1K_2)
e^{-2\frac{\xi}{c}K_1d}}\right]^2
e^{-2\frac{\xi}{c}pa},
\nonumber\\
&&
Q_2(i\xi)=1-\left[\frac{(K_1-p)(K_1+K_2)-(K_1+p)(K_1-K_2)
e^{-2\frac{\xi}{c}K_1d}}{(K_1+p)(K_1+K_2)-(K_1-p)(K_1-K_2)
e^{-2\frac{\xi}{c}K_1d}}\right]^2
e^{-2\frac{\xi}{c}pa}.
\label{17}
\ees
\noindent}
Here all permittivities depend on $i\xi$ and
\beq
K_{\alpha}=K_{\alpha}(i\xi)\equiv\sqrt{p^2-1+\varepsilon_{\alpha}(i\xi)}=
\frac{c}{\xi}R_{\alpha}(i\xi),
\qquad
 \alpha=1,\,2.
\label{18}
\eeq
\noindent
For $\alpha=0$ one has $p=cR_0/\xi$ which is equivalent to (\ref{15}).

Notice that the expressions (\ref{13}), (\ref{16}) give us the finite values
of the Casimir energy density which is in less common use than the force.
Thus in \cite{5} no finite expression for the energy density is presented
for two semi-spaces. In \cite{27} the omission of infinities is performed
implicitly, namely instead of Eqs.~(\ref{3}) the result of their division
by the terms containing $\exp(R_0a)$ was presented. The coefficient near
$\exp(R_0a)$, however, turns into infinity on $C_{+}$. In other words the
Eqs.~(\ref{3}) are divided by infinity. As a result the integral along
$C_{+}$ is equal to zero in \cite{27} and the quantity (\ref{2}) would
seem to be finite. Fortunately, this implicit division is equivalent to the
regularization procedure explicitly presented above. That is why the final
results obtained in \cite{27} are indeed correct. In \cite{11} the energy 
density is not considered at all.

From (\ref{16}) it is easy to obtain the Casimir force 
per unit area acting between
semi-spaces covered with layers
\beq
F_{ss}(a,d)=-\frac{\partial{\cal E}_{reg}(a,d)}{\partial a}=
-\frac{\hbar}{2\pi^2c^3}
\int\limits_{1}^{\infty} p^2\,dp
\int\limits_{0}^{\infty} \xi^3\,d\xi
\left[\frac{1-Q_1(i\xi)}{Q_1(i\xi)}+ 
\frac{1-Q_2(i\xi)}{Q_2(i\xi)}\right].
\label{19}
\eeq
\noindent
This expression coincides with Lifshitz result [9--11] for the force 
per unit area between
semi-spaces with a dielectric permittivity $\varepsilon_2$ if the covering
layers are absent. To obtain this limiting case from (\ref{19}) one should
put $d=0$ and $\varepsilon_1=\varepsilon_2$
\beq
F_{ss}(a)=
-\frac{\hbar}{2\pi^2c^3}
\int\limits_{1}^{\infty} p^2\,dp
\int\limits_{0}^{\infty} \xi^3\,d\xi
\left\{\left[\left(\frac{K_2+\varepsilon_2p}{K_2-\varepsilon_2p}\right)^2
e^{2\frac{\xi}{c}pa}-1\right]^{-1}
+\left[\left(\frac{K_2+p}{K_2-p}\right)^2
e^{2\frac{\xi}{c}pa}-1\right]^{-1}\right\}.
\label{20}
\eeq
\noindent
The corresponding quantity for the energy density follows from (\ref{16})
\beq
{\cal E}_{reg}(a)=
\frac{\hbar}{4\pi^2c^2}
\int\limits_{1}^{\infty} p\,dp
\int\limits_{0}^{\infty} \xi^2\,d\xi
\left\{\ln\left[1-\left(\frac{K_2-\varepsilon_2p}{K_2+\varepsilon_2p}\right)^2
e^{-2\frac{\xi}{c}pa}\right]
+\ln\left[1-\left(\frac{K_2-p}{K_2+p}\right)^2
e^{-2\frac{\xi}{c}pa}\right]\right\}.
\label{21}
\eeq 

The other possibility to obtain the force between semi-spaces (but with 
a permittivity
$\varepsilon_1$) is to consider limit $d\to\infty$ in (\ref{19}).
In this limit we obtain once more the results (\ref{20}), (\ref{21})
where $K_2$, $\varepsilon_2$ are replaced by $K_1$, $\varepsilon_1$. Note
also that we do not take into account the effect of non-zero point temperature
which is negligible for $a\ll\hbar c/T$.

The independent expression for the physical energy density is especially
important because it allows the possibility to obtain approximate value of
the force for the configuration of a sphere (or a spherical lens) above a
semi-space. Both bodies can be covered by the layers of another material.
According to the proximity force theorem this force is
\beq
F_{sl}(a,d)=2\pi R
{\cal E}_{reg}(a,d)=\frac{\hbar R}{2\pi c^2}
\int\limits_{1}^{\infty} p\,dp
\int\limits_{0}^{\infty} \xi^2\,d\xi
\left[\ln Q_1(i\xi)+ \ln Q_2(i\xi)\right],
\label{22}
\eeq
\noindent
where $R$ is the sphere radius, $Q_{1,2}$ are defined in (\ref{17}). 
In the absence of layers ${\cal E}_{reg}(a,d)$ should be substituted by
${\cal E}_{reg}(a)$ from (\ref{21}).

Although the expression (\ref{22}) is not exact it allows the possibility
to calculate the force with a very high accuracy. As was shown in \cite{6}
(see also \cite{31,32}) the proximity force theorem is equivalent to
additive summation of interatomic van der Waals and Casimir force
potentials with a subsequent normalization of the interaction constant.
As was shown in \cite{33} the accuracy of such method is very high
(the relative error of the obtained results is less than 0.01\%) if the
configuration corresponds closely with two semi-spaces which is the case
for a sphere (lens) of a large radius $R\gg a$ above a semi-space.

In the following Sections the above general results will be used for
computation of the Casimir and van der Waals forces acting between real
metals.

\section{THE INFLUENCE OF FINITE CONDUCTIVITY ON THE CASIMIR FORCE}

Let us first consider semi-spaces made of aluminum or gold. Aluminum
covered interacting bodies (a plate and a lens) were used in the
experiments \cite{4,8} because of its high reflectivity for wavelengths
(plate-sphere separations) larger than 100\,nm. The thickness of $Al$
covering layer was 300\,nm. It is significantly greater than the effective
penetration depth of the electromagnetic zero point oscillations into $Al$ 
which is
$\delta_0=\lambda_p/(2\pi)\approx 17\,$nm (see Introduction). That is why
$Al$ layer can be considered as infinitely thick and modelled by a semi-space.
In the experiment \cite{3} the test bodies were covered by a 500\,nm $Au$
layer which also can be considered as infinitely thick. In \cite{4} and
\cite{8} $Al$ surfaces were covered, respectively, by $d<20\,$nm and
$d=8\,$nm sputtered $Au/Pd$ layers to reduce the oxidation processes in $Al$
and the effect of any associated electrostatic charges.  The influence of such 
additional thin layers on the Casimir force is
discussed in Sec.~IV.

The values of the force per unit area
for the configuration of two semi-spaces and the force for a sphere 
above a semi-space are given by Eq.~(\ref{20}) and Eqs.~(\ref{21}), (\ref{22}).
For the distance $a$ much larger than the characteristic wavelength of
absorption spectra of the semi-space material $\lambda_0$
Eqs.~(\ref{20}), (\ref{21}) lead \cite{11} to the following results in the
case of ideal metal ($\varepsilon_2\to\infty$)
\beq
F_{ss}^{(0)}(a)=-\frac{\pi^2}{240}\frac{\hbar c}{a^4}, \qquad
F_{sl}^{(0)}(a)=-\frac{\pi^3}{360}R\frac{\hbar c}{a^3}.
\label{23}
\eeq

To calculate numerically the corrections to (\ref{23}) due to the finite
conductivity of a metal we use the tabulated data for the complex index
of refraction $n+ik$ as a function of frequency \cite{34}. The values of
dielectric permittivity along the imaginary axes can be expressed through 
Im$\varepsilon(\omega)=2nk$ with the help of dispersion relation \cite{11}
\beq
\varepsilon(i\xi)=1+\frac{2}{\pi}
\int\limits_{0}^{\infty}
\frac{\omega\,\mbox{Im}\varepsilon(\omega)}{\omega^2+\xi^2}d\omega.
\label{24}
\eeq
\noindent

Here the complete tabulated refractive indices extending from 0.04\,eV 
to 10000\,eV for Al and from 0.1\,eV to 10000\,eV for Au from \cite {34} are 
used to calculate Im$\varepsilon(\omega)$. For frequencies below 0.04 eV 
in the case of Al and below 0.1 eV in the case of Au, the table values 
of \cite {34} can be extrapolated using the free electron Drude model. 
In this case, the dielectric
permittivity along the imaginary axis is represented as:
\beq
\varepsilon_{\alpha}(i\xi)=1+\frac{\omega_{p\alpha}^2}{\xi(\xi+\gamma)},
\label{25}
\eeq
\noindent
where $\omega_{p\alpha}=(2\pi c)/\lambda_{p\alpha}$ is the plasma frequency 
and $\gamma$ is the relaxation frequency.  A $\omega_{p}$=12.5\,eV and 
$\gamma$=0.063\,eV was used for the case of Al based on the last 
results in Table XI 
on p.394 of 
\cite {34}. In the case of Au the analysis is not as straightforward, 
but proceeding in the manner outlined in \cite {25} we obtain 
$\omega_{p}$=9.0\,eV and $\gamma$=0.035\,eV. While the values of $\omega_{p}$ 
and $\gamma$ based on the optical data of various sources might differ 
slightly we have found that the resulting numerically computed Casimir 
forces to differ by less than 1\%. In fact, if for Al metal, a 
$\omega_p$=11.5\,eV and $\gamma$=0.05\,eV
as in \cite{25} is used, the differences are extremely small. Of the values 
tabulated 
below, only the value of the 
force in the case of a sphere and a semi-space at 0.5\,$\mu$m separation is 
increased by 0.1\% which on round-off to the second significant figure 
leads to an increase of 1\%.
The results of numerical integration 
by Eq.~(\ref{24}) for $Al$ (solid
curve) and $Au$ (dashed curve) are presented in Fig.2 in a logarithmic
scale. As is seen from Fig~2 the dielectric permittivity along the
imaginary axis decreases monotonically with increasing frequency (in
distinction to Im$\varepsilon(\omega)$ which possesses peaks corresponding
to inter-band absorption).

The obtained values of the dielectric permittivity along the imaginary axis 
were substituted into Eqs.~(\ref{20}) and (\ref{22}) (with account of
(\ref{21})) to calculate the Casimir force acting between real metals in
configurations of two semi-spaces (ss) and a sphere (lens) above 
a semi-space (sl).  Numerical integration was done from an upper limit 
of $10^{4}\,$eV to a lower limit of $10^{-6}\,$eV.  Changes in the 
upper limit or 
lower limit by a factor of 10 lead to changes of less than 0.25\% in 
the Casimir force.  If the trapezoidal rule is used in the numerical 
integration of Eqs.~(\ref{24}) the corresponding Casimir force decreases 
by a factor less than 0.5\%.   The results are presented in Fig.~3(a) 
(two semi-spaces) 
and in Fig.~3(b) for a sphere
above a semi-space by the solid lines 1 (material of the test bodies is
aluminum) and 2 (material is gold). In the vertical axis the relative
force $F_{ss}/F_{ss}^{(0)}$ is plotted in Fig.~3(a) and $F_{sl}/F_{sl}^{(0)}$
in Fig.~3(b). 
These quantities provide a sense of the correction factors to the Casimir
force due to the effect of finite conductivity. In the horizontal axis
the space separation is plotted in the range 0.1--1{\mum}. We do not
present the results for larger distances because the temperature
corrections to the Casimir force become significant.
 At room temperature
the temperature corrections contribute only 2.6\% of $F_{sl}^{(0)}$ at
$a=1\,\mu$m, but at $a=3\,\mu$m they contribute 47\% of $F_{sl}^{(0)}$,
 and
at $a=5\,\mu$m --- 129\% of $F_{sl}^{(0)}$ \cite{35}. It is seen that the
relative force for $Al$ is larger than for $Au$ at the same separations
as it should be because of better reflectivity properties of $Al$.

It is interesting to compare the obtained results with those of
Refs.~\cite{23,24} and \cite{25,26} where the similar computations were
performed (in \cite{25,26} the analytical expressions equivalent to
Eqs.~(\ref{20}) and (\ref{21}) were used, in \cite{23,24}, however,
the energy density between plates was obtained by a numerical integration
of the force which can lead to some additional error). All the results for
the several values of distance between the test bodies are presented in the
Table 1.
\begin{table}[ht]
\caption{The correction factor to the Casimir force due to the 
finite conductivity of the metal by the results of different authors
and the present paper in configurations of two semi-spaces (ss) and
a sphere (lens) above a semi-space (sl).}
\begin{tabular}{ccccccc} 
 Test & Metal & $a$ & \multicolumn{4}{c}{$F/F^{(0)}$}
 \\ 
bodies&&($\mu$m)& \multicolumn{3}{c}{Computation}  & Perturbation \\
&&&\cite{23,24} & \cite{25,26} & This paper
& theory \cite{17}\\ \hline
ss & $Al$ & 0.1 & 0.557 & 0.55 & 0.55 & 0.56 \\ 
sl & $Al$ & 0.1 & 0.651 & 0.63 & 0.62 & 0.61 \\ 
ss & $Au$ & 0.1 & --- & 0.48 & 0.49 & 0.62 \\ 
sl & $Au$ & 0.1 & --- & 0.55 & 0.56 & 0.60 \\ 
ss & $Al$ & 0.5 & --- & 0.85 & 0.84 & 0.84 \\ 
sl & $Al$ & 0.5 & --- & 0.88 & 0.87 & 0.88 \\ 
ss & $Au$ & 0.5 & 0.657 & 0.81 & 0.81 & 0.81 \\ 
sl & $Au$ & 0.5 & 0.719 & 0.85 & 0.85 & 0.85 \\ 
sl & $Au$ & 0.6 & 0.78 & 0.87 & 0.87 & 0.87 \\ 
ss & $Al$ & 3 & --- & 0.96 & 0.96 & 0.97 \\ 
sl & $Al$ & 3 & --- & 0.97 & 0.97 & 0.98 \\ 
ss & $Au$ & 3 & --- & 0.96 & 0.95 & 0.96 \\ 
sl & $Au$ & 3 & --- & 0.97 & 0.96 & 0.97 
\end{tabular}
\end{table}
As is seen from Table 1, our calculational results (column 6) are in
agreement with \cite{25,26} (column 5) up to 0.01. At the same time the
results of \cite{23,24} (column 4) for $Au$ are in disagreement with
both \cite{25,26} and this paper. The results for $Al$ are presented in
\cite{23} at $a=0.1\,\mu$m only. Note that the results at $a=3\,\mu$m (the
last four lines of the Table 1) are valid only at zero temperature. 
They do 
not take into account temperature corrections which are significant for
such separation. Also the results of \cite{23,24} for $Cu$ covered bodies
are in disagreement with \cite{25,26}. We do not consider $Cu$ here
because the outer surfaces in the recent experiments were covered by the
thick layers of $Au$ \cite{3} and $Al$ \cite{4,8}. The hypothesis of
\cite{24} that the $Au$ film of 0.5{\mum} thickness could significantly
diffuse into the $Cu$ layer of the same thickness at room temperatures 
seems unlikely.
In any case it is not needed because the dielectric permittivity of $Au$
and $Cu$ along the imaginary axis is almost the same \cite{17,25,26} and,
consequently, will also lead to the same Casimir force.

The computational results obtained here are in good agreement with analytical
perturbation expansions of the Casimir force in powers of relative
penetration depth $\delta_0=\lambda_p/(2\pi)$ of the electromagnetic zero point
oscillations into the metal. 
Representation (\ref{25}) with $\gamma=0$ 
is applicable for the wavelengths (space
separations) larger than $\lambda_{p\alpha}$ (the corrections due to
relaxation processes are small for the distances $a\leq 5\,\mu$m).
It can be substituted into Eqs.~(\ref{20}), (\ref{21}) to get the
perturbation expansion. According to the results of Ref.~\cite{17} the
relative Casimir force with finite conductivity corrections up to the 4th
power is
\beq
\frac{F_{ss}(a)}{F_{ss}^{(0)}(a)}=1-\frac{16}{3}\frac{\delta_0}{a}+
24\frac{\delta_0^2}{a^2}-
\frac{640}{7}\left(1-\frac{\pi^2}{210}\right)\frac{\delta_0^3}{a^3}
+
\frac{2800}{9}\left(1-\frac{163\pi^2}{7350}\right)\frac{\delta_0^4}{a^4}
\label{26}
\eeq
\noindent
for two semi-spaces and
\beq
\frac{F_{sl}(a)}{F_{sl}^{(0)}(a)}=1-4\frac{\delta_0}{a}+
\frac{72}{5}\frac{\delta_0^2}{a^2}-
\frac{320}{7}\left(1-\frac{\pi^2}{210}\right)\frac{\delta_0^3}{a^3}
+
\frac{400}{3}\left(1-\frac{163\pi^2}{7350}\right)\frac{\delta_0^4}{a^4}
\label{27}
\eeq
\noindent
for a sphere (lens) above a semi-space.

In Fig.~3(a) (two semi-spaces) the dashed line 1 represents the results
obtained by (\ref{26}) for $Al$ with $\lambda_p=107\,$nm
(which corresponds to $\omega_p=11.5\,$eV), and the dashed
line 2 --- the results obtained by (\ref{26}) for $Au$ with
$\lambda_p=136\,$nm ($\omega_p=9\,$eV) \cite{25}. 
In Fig.~3(b) the dashed lines 1 and 2 
represent the
perturbation results obtained for $Al$ and $Au$ by (\ref{27}) for a lens
above a semi-space. As is seen from the last column of Table 1,
the perturbation results are in good (up to 0.01) agreement with
computations for all distances larger than $\lambda_p$. Only at
$a=0.1\,\mu$m for $Au$ there are larger deviations because
$\lambda_{p1}\equiv\lambda_p^{Au}>0.1\,\mu$m.

\section{THE CASIMIR FORCE BETWEEN LAYERED \protect{\\} SURFACES}

In this Section we consider the influence of the thin outer metallic layers
on the Casimir force value. Let the semi-space made of $Al$ ($\varepsilon_2$)
be covered by $Au$ ($\varepsilon_1$) layers as shown in Fig.~1.
For a configuration of a sphere above a plate such covering made of $Au/Pd$
was used in experiments \cite{4,8} with different values of layer
thickness $d$. In this case the Casimir force is given by the Eqs.~(\ref{19}),
(\ref{22}), where the quantities $Q_{1,2}(i\xi)$ are expressed by 
Eqs.~(\ref{17}), (\ref{18}). The computational results for
$\varepsilon_{\alpha}(i\xi)$ are obtained in the previous Section by
Eq.~(\ref{24}). Substituting them into (\ref{19}), (\ref{22}) and
performing  a numerical integration in the same way as above
one obtains the Casimir force including the effect of covering layers. 
The computational
results for a configuration of two semi-spaces are shown in Fig.~4(a). 
Here the 
solid lines represent once more the Casimir force between semi-spaces of 
pure $Al$ and $Au$ respectively, the dashed and dotted lines 
are for the case of
$Au$ layers of thickness $d=20\,$nm and $d=30\,$nm covering $Al$. When the
layers are present, the space separation $a$ is measured from their outer
surfaces according to Eqs.~(\ref{19}), (\ref{22}). In Fig.~4(b) the analogous
results with the same notations are presented for the configuration of a
sphere (lens) above a semi-space.

As is seen from Fig.~4, the $Au$ layer of $d=20\,$nm thickness
significantly decreases the relative Casimir force between $Al$ surfaces.
With this layer the force approaches the value for pure $Au$
semi-spaces. For a thicker $Au$ layer of $d=30\,$nm thickness the relative
Casimir force is scarcely affected by the underlying $Al$. For example, 
at a space
separation $a=300\,$nm in the configuration of two semi-spaces we have
$F_{ss}/F_{ss}^{(0)}=0.773$ for pure $Al$,
$F_{ss}/F_{ss}^{(0)}=0.727$ for $Al$ with 20\,nm $Au$ layer,
$F_{ss}/F_{ss}^{(0)}=0.723$ for $Al$ with 30\,nm $Au$ layer,
and $F_{ss}/F_{ss}^{(0)}=0.720$ for pure $Au$.
In the same way for the configuration of a sphere above a semi-space 
the results
are:
$F_{sl}/F_{sl}^{(0)}=0.817$ (pure $Al$),
0.780 ($Al$ with 20\,nm $Au$ layer),
0.776 ($Al$ with 30\,nm $Au$ layer),
0.774 (pure $Au$). Both limiting cases $d\to\infty$ and $d\to 0$ were
considered and the results are shown to coincide with that of Sec.~III.

Let us now discuss the application range of the obtained results for 
the case of covering layers. First from a theoretical standpoint, the main
question concerns the layer thicknesses to which the obtained formulas 
(\ref{19}),
(\ref{22}) and the above computations can be applied. In the derivation of
Sec.~II the spatial dispersion is neglected and, as a consequence, the
dielectric permittivities $\varepsilon_{\alpha}$ depend only on $\omega$
not on the wave vector {\boldmath$k$}. In other words the field of vacuum
oscillations is considered as time-dependent but space homogeneous. Except
for the thickness of a skin layer $\delta_0$ the main parameters of our problem
are the velocity of the electrons on the Fermi surface $v_F$, the
characteristic frequency of the oscillation field $\omega$, and the mean free 
path of the electrons $l$. For the considered region of high
frequencies (micrometer distances between the test bodies) the following
conditions are valid \cite{36}
\beq
\frac{v_F}{\omega}<\delta_0\ll l.
\label{28}
\eeq
\noindent
Note that the quantity $v_F/\omega$ in the left-hand side of Eq.~(\ref{28})
is the distance travelled by an electron during one period of the field, so
that the first inequality is equivalent to the assumption of spatial
homogeneity of the oscillating field. Usually the corresponding frequencies
start from the far infrared part of spectrum which means the space
separation $a\sim 100\,\mu$m \cite{6}. The region of high frequencies is
restricted by the short-wave optical or near ultraviolet parts of the spectrum
which correspond to the surface separations of several hundred nanometers.
For smaller distances absorption bands, photoelectric effect and other physical
phenomena should be taken into account. For these phenomena, the general 
Eqs.~(\ref{19}), 
(\ref{22}), however, are still valid if one substitutes the experimental
tabulated data for the dielectric permittivity along the imaginary axis
incorporating all these phenomena.

Now let us include one more physical parameter --- the thickness $d$ of the
additional, i.e. $Au$, covering layer.
It is evident that Eqs.~(\ref{19}), (\ref{22}) are applicable only for
layers of such thickness that
\beq
\frac{v_F}{\omega}<d.
\label{29}
\eeq
\noindent
Otherwise an electron goes out of the thin layer during one period
of the oscillating field and the approximation of space homogeneity is 
not valid. If $d$ is so small that the inequality (\ref{29}) is violated 
the spatial
dispersion should be taken into account which means that the dielectric
permittivity would depend not only on frequency but on a wave vector also:
$\varepsilon_1=\varepsilon_1(\omega,{\mbox{\boldmath$k$}})$. So, if (\ref{29})
is violated the situation is analogous to the anomalous skin effect where
only space dispersion is important and the inequalities below are valid
\beq
\delta_0(\omega)<\frac{v_F}{\omega}, 
\qquad \delta_0(\omega)<l.
\label{30}
\eeq
\noindent
In our case, however, the role of $\delta_0$ is played by the layer
thickness $d$ (the influence of nonlocality effects on van der Waals force
is discussed in \cite{37,37a}).

From (\ref{28}), (\ref{29}) it follows that for pure $Au$ layers
($\lambda_p\approx 136\,$nm) the space dispersion can be neglected only if
$d\geq(25-30)\,$nm. For thinner layers a more general theory taking into
account nonlocal effects should be developed to calculate the Casimir force.
Thus for such thin layers the bulk tabulated data of the dielectric 
permittivity
depending only on frequency cannot be used (see experimental investigation
\cite{38} demonstrating that for $Au$ the bulk values of dielectric
constants can only be obtained from films whose thickness is about 30\,nm
or more). That is why the dashed lines  in Fig.~4 ($d=20\,$nm layers)
are subject to corrections due to the influence of spatial dispersion,
whereas the solid lines  represent the final result. From an experimental 
standpoint thin layers of order a few nm grown by evaporation or sputtering 
techniques are 
highly porous. This is particularly so in the case of sputtered 
coatings as shown in \cite{38e}. The nature of porosity is a function of the 
material and 
the underlying substrate. Thus it should be noted that the theory presented 
here which 
used the bulk tabulated data for
$\varepsilon_1$ cannot be applied to calculate the influence of thin
covering layers of $d<20\,$nm \cite{4,7} and of $d=8\,$nm \cite{8,39}
on the Casimir force. The measured high transparency of such layers for the
characteristic frequencies \cite{4,7} corresponds to a larger change of 
the force
than what follows from the Eqs.~(\ref{19}), (\ref{22}). This is in agreement
with the above qualitative analyses.

The role of spatial dispersion was also neglected in the paper \cite{38a} 
where an
attempt was made to describe theoretically the influence of thin 
metallic covering layers onto the Casimir force in
experiments \cite{4,8}. Also the bulk materials properties were used for the 
$Au/Pd$ films. As shown in \cite{38d}, the resistivity of 
sputtered $Au/Pd$ films even
of 60nm thickness have been shown to be extremely high of order 
2000\,ohm$\cdot$cm.   In \cite{38a} it was concluded that the
maximum possible theoretical values of the force including the covering 
layers is
significantly smaller than the measured ones. The data of \cite{4,8} 
is however shown to be consistent with a theory neglecting the influence 
of layers. In \cite{4,8} the surface separations are calculated from $Al$ 
surfaces. Including the thickness of covering layers reduces the distance 
between the outer 
surfaces which is now smaller than the distance between $Al$ surfaces. 
Thus contrary to \cite{38a}, the theoretical value of force should 
increase when the 
presence of the layers is included. The error made in \cite{38a} 
can be traced to the following. The authors of \cite{38a} changed the 
data of \cite{8} 
``by shifting all the
points to larger separations on $2h=16\,$nm'' (where $h=8\,$nm is the
layer thickness in \cite{8}) instead of shifting to smaller separations 
by 16nm as based on \cite{8}. If the correct shift is done then the 
theoretical 
values of the force, including  the effect of covering layers, 
are not smaller than the experimental values.
Hence the conclusion
in \cite{38a} about the probable influence of new hypothetical 
attractions based on
the experiments \cite{4,8} is unsubstantiated.

\section{THE VAN DER WAALS FORCE AND INTERMEDIATE REGION}

As is seen from Figs.~3,4 at room temperature the Casimir force does not
follow its ideal field-theoretical expressions (\ref{23}).
 For the
space separations less than $a=1\,\mu$m the corrections due to finite
conductivity of the metal are rather large 
(thus, at $a=1\,\mu$m they are
around 7--9\% for a lens above a semi-space, and 10--12\% for two
semi-spaces; at $a=0.1\,\mu$m --- around 38--44\% (sl), and 45--52\% (ss)).
For $a>1\,\mu$m the temperature corrections increase very quickly (see
Sec.~III). Actually, the range presented in Figs.~3,4 is the beginning 
of a transition with decreasing $a$ from the Casimir force to the van 
der Waals force. 
Our aim is to investigate the intermediate region in more
detail for smaller $a$ and to find values of $a$ where the pure
(non-retarded) van der Waals regime starts. To do this for the case when no
additional covering layers are present we numerically evaluate the
integrals in Eqs.~(\ref{20})--(\ref{22}) for $a<100\,$nm.

The computational results obtained by the same procedures as in Sec.~III
are presented in Fig.~5(a) for two semi-spaces  and 5(b) for a sphere above 
a semi-space.
In both Figs. the solid line represents the results for aluminum test
bodies, and the dashed line for gold ones. The absolute values of the van
der Waals force and surface separation $a$ are plotted along the vertical and
horizontal axes in a logarithmic scale. 
The asymptotic expressions in the limit of
$a\ll\lambda_0$ following from
Eqs.~(\ref{20})--(\ref{22}) respectively are 
\cite{11}
\beq
F_{ss}^{(0)}(a)=-\frac{H}{6\pi a^3}, \qquad
F_{sl}^{(0)}(a)=-\frac{HR}{6a^2}.
\label{31}
\eeq
\noindent
Here it is important to note that
the Hamaker constant $H$ is dependent on the material properties of
the boundaries and is 
a priori unknown.
This is in contrast to the ideal Casimir force limit of Eq.~(\ref{23})
(obtained for $a\gg\lambda_0$) which is material independent and is only
a function of $\hbar$ and $c$. Thus it is not reasonable to express the
van der Waals force as a ratio relative to Eq.~(\ref{31}).
The asymptotic behavior (\ref{31}) will be used below
to determine the value of  $H$.

The computations were performed with a step $\Delta a=5\,$nm in the interval
10\,nm$\leq a\leq 100\,$nm, $\Delta a=1\,$nm in the interval
4\,nm$\leq a\leq 10\,$nm, $\Delta a=0.2\,$nm in the interval
2\,nm$\leq a\leq 4\,$nm, and $\Delta a=0.1\,$nm for
0.5\,nm$\leq a\leq 2\,$nm. 
At $a=100\,$nm the force values coincide with those in Fig.~3.
For $a<0.5\,$nm the repulsive exchange 
forces dominate. As is seen from Fig.~5 for both configurations and two 
metals under consideration ($Al$ and $Au$) the range of purely van der
Waals force described by Eqs.~(\ref{31}) turn out to be extremely
narrow. It extends from 0.5\,nm till 2--4\,nm only. For larger distances
the transition from the force-distance dependence $\sim a^{-3}$ to the
dependence $\sim a^{-4}$ begins (for two semi-spaces) and from the
dependence $\sim a^{-2}$ to $\sim a^{-3}$ (for a lens above a semi-space).
This conclusion is in a qualitative agreement with the results of \cite{18}
where the van der Waals force between a metallic sample and a metallic tip
of the atomic force microscope was calculated (our choice of a sphere is
formally equivalent to the paraboloidal tip considered in \cite{18}).
Calculation in \cite{18} was performed by numerical integration of
Lifshitz-type equation for the force with the permittivity of a metal
given by the plasma model [Eq.(\ref{25}) with $\gamma=0$]. 
Strictly speaking plasma model
is not applicable for $a\ll\lambda_0$ (see Sec.~III). That is why we have 
used the optical tabulated data for the complex refractive index in
our computations. However, the correct conclusion about the extremely
narrow distance range of the purely van der Waals region for metals 
is obtainable by using the plasma model to represent their dielectric 
properties. Note that for dielectric test bodies the pure 
van der Waals regime extends to larger distances. For example in the
configuration of two crossed mica cylinders (which is formally equivalent to
a sphere above a semi-space) the van der Waals regime extends from 1.4\,nm
till 12\,nm as was experimentally shown in \cite{40}.

\section{DETERMINATION OF HAMAKER CONSTANTS FOR $Al$ AND $Au$}

The results of the previous Sec. make it possible to determine the
values of the Hamaker constant $H$ from Eq.(\ref{31}) for aluminum and
gold. Let us start with the configuration of two semi-spaces. As is seen from
the computational results presented in Fig.~5(a) (solid curve) 
the asymptotic regime
for $Al$ extends here from $a=0.5\,$nm till $a=4\,$nm. 
We use a more narrow interval 0.5\,nm--2\,nm for the determination of $n$
and $H$. The power index $n$
of the force-distance relation given by the first formula of Eq.~(\ref{31})
is equal to $n=3.02\pm 0.01$ in the considered interval. 
To obtain this value the slopes between adjacent points, i.e. (0.5--0.6)\,nm,
(0.6--0.7)\,nm etc were calculated and then the average and the standard
deviation were found.
The corresponding mean
value of the Hamaker constant is
\beq
H_{ss}^{Al}=(3.67\pm 0.02)\times 10^{-19}\,\mbox{J}.
\label{32}
\eeq

Considering the computational results for $Au$ 
(dashed curve of Fig.~5(a)) we find
the asymptotic regime in a more narrow interval 0.5\,nm\,--\,2\,nm with the
power index $n=3.04\pm 0.02$. The mean value of the Hamaker constant
turns out to be equal to
\beq
H_{ss}^{Au}=(4.49\pm 0.07)\times 10^{-19}\,\mbox{J}.
\label{33}
\eeq

For the configuration of a sphere (lens) above a semi-space the results are
presented in Fig.~5(b) (solid curve for $Al$ and dashed curve for $Au$). 
In both cases
the asymptotic region extends from $a=0.5\,$nm till $a=2\,$nm only with
the mean values of power index in the second formula of Eq.~(\ref{31})
$n=2.04\pm 0.02$ ($Al$) and $n=2.08\pm 0.03$ ($Au$). The corresponding
mean values of the Hamaker constant are
\beq
H_{sl}^{Al}=(3.60\pm 0.06)\times 10^{-19}\,\mbox{J},
\qquad
H_{sl}^{Au}=(4.31\pm 0.14)\times 10^{-19}\,\mbox{J}.
\label{34}
\eeq
\noindent
It is seen that in the case of $Au$ and a sphere above a semi-space
configuration the behavior of the force shows less precise 
agreement with the second
formula of Eq.~(\ref{31}).

The above results obtained for the two configurations independently give the
possibility to derive new values of the Hamaker constant for $Al$
and $Au$. Taking into account the value of (\ref{32}) and the first
expression from (\ref{34}) we get
\beq
H^{Al}=(3.6\pm 0.1)\times 10^{-19}\,\mbox{J}.
\label{35}
\eeq
\noindent
The absolute error here was chosen in such a way to cover both permitted 
intervals in (\ref{32}) and (\ref{34}).

For $Au$ the tolerances of the second value from (\ref{34}) 
are two times wider than
the permitted interval from (\ref{33}). That is why the most probable final
value of the Hamaker constant for gold can be estimated as
\beq
H^{Au}=(4.4\pm 0.2)\times 10^{-19}\,\mbox{J}.
\label{36}
\eeq
\noindent
The decreased accuracy than in (\ref{35}) is explained by the extremely narrow 
region of pure van der Waals force law for gold.  These values of $H$ 
for gold are compatible with those obtained previously. 
For example, in \cite{45} values 
between 
$(2-4) \times 10^{-19}\,${J} were obtained using different procedures.

\section{CONCLUSIONS AND DISCUSSION}

In the above, general expressions were obtained both for the Casimir
energy density and force in the configuration of two 
plates (semi-spaces) with different separations between them. The case of 
where the surfaces were covered by the thin layers made of the another 
material was also considered.  
Additional clarifications of the regularization procedure were given.
This is important for obtaining a finite physical value for the energy
density. The latter quantity is very important for obtaining the
Casimir force for the configuration of a sphere (lens) above 
a plate (semi-space) which 
was used in the recent experiments. For this configuration the general
expression for the Casimir force with account of layers covering a lens
and a semi-space was arrived at by the use of proximity force theorem.

The Casimir force was recalculated between $Al$ and $Au$ test bodies for the
configurations of two semi-spaces and a sphere (lens) above a semi-space.
The disagreement between the results of \cite{23,24} and \cite{25,26} was
resolved in favor of \cite{25,26}. 
Additionally, computational results were
compared with perturbation expansion up to the fourth order in powers of
relative penetration depth of electromagnetic zero point oscillations into the
metal. The perturbation results are also in agreement
with \cite{25,26} and our computations for the space separations larger
than a plasma wavelength of the metal under study (not much larger as it
to be expected from general considerations). 
We have performed the first computations 
of the Casimir
force between $Al$ test bodies covered by $Au$ thin layers. The monotonous
decrease of the correction factor to the Casimir force was observed with 
increase of the layer thickness. The qualitative analysis leads to
the conclusion that the thickness of the layer should be large enough to 
allow neglect of the spatial dispersion of the dielectric permittivity
and the use of bulk optical tabulated data for the complex refractive index.
For the $Au$ layers the minimal allowed thickness for such an approximation 
was estimated as
$d=30\,$nm in agreement with the experimental evidence of \cite{38}. For
smaller layer thicknesses the bulk optical tabulated data cannot be used.
In this case the calculation of the Casimir force would require a direct
measurement of the complex refractive index for the particular metal (not
only the frequency dependence but also its dependence
on the wave vector).

The van der Waals force was calculated between the $Al$ and $Au$ test bodies
in configurations of two semi-spaces and a sphere (lens) above a semi-space.
The computations were performed starting from the same general expressions
as in the case of the Casimir force and using the same numerical procedure
and optical tabulated data. The extremely
narrow region where the pure non-retarded van der Waals
power-law force acts was noted. This region extends from $a=0.5\,$nm till
$a=(2-4)\,$nm only. For larger distances a wide transition region starts, 
where the non-retarded van der Waals force
described by the Eq.~(\ref{31}) gradually transforms into the retarded
van der Waals (Casimir) force from the Eq.~(\ref{23}) when the space
separation approaches the value $a=1\,\mu$m. The values of the Casimir
force given by the Eq.~(\ref{31}) are never achieved at room temperature
(at $a=1\,\mu$m due to the finite conductivity of the
metal while for larger distances the temperature corrections
make a strong contribution). Using the asymptotic region of the pure
non-retarded van der Waals force the new values of the Hamaker constant for
$Al$ and $Au$ were obtained. For $Al$ the reported accuracy corresponds to a
relative error of 2.8\%, and for $Au$ it is around 4.5\%.

The obtained results do not exhaust all the problems connected with the
role of finite conductivity of the metal in the precision measurements
of the Casimir force. The main problem to be solved is the investigation
of corrections to the force due to thin covering layers. This would demand
theoretical work on the generalization of the Lifshitz formalism for the
case when the spatial dispersion can be important in addition to the 
frequency dependence.
Also the new measurements of the complex refractive index are needed
for the layers under consideration. What's more the finite conductivity
corrections to the Casimir force should be considered together with the
corrections due to the surface roughness (see, e.g., \cite{7} where the
non-additivity of both influential factors is demonstrated) and corrections 
due to finite temperature. This combined research is necessary for both 
applied and fundamental applications of the Casimir effect. It has been known 
that the measurements of the Casimir force give the possibility to obtain 
strong constraints for the constants of long-range interactions and light
elementary particles predicted by the unified gauge theories, supersymmetry
and supergravity \cite{6}. Such information is unique and cannot be obtained
even by means of the most powerful modern accelerators. In Ref.~\cite{35} the
constraints for the Yukawa-type hypothetical interactions were strengthened 
up to
30 times in some distance range on the base of Casimir force measurements
of Ref.~\cite{3}. The increased precision of the Casimir force in \cite{4} 
gave the possibility to strengthen constraints up to 140 times on the
Yukawa-type interactions at smaller distances \cite{41}.
It is highly probable that the new measurements of the Casimir force with
increased accuracy will serve as an important alternative source of 
information about the elementary particles and fundamental interactions.

\section*{ACKNOWLEDGMENTS}

G.L.K. and V.M.M. are grateful to the Department of Physics of the
Federal University of Paraiba, where this work was partly done,
for their hospitality.

\newpage
\begin{center} {\Large {\bf List of captions}} \end{center}
\begin{tabular}{lcp{142mm}}
& &\\
{\bf FIG.1.}& & 
The configuration of two semi-spaces with a dielectric permittivity
$\varepsilon_2(\omega)$ covered by layers of thickness $d$
with a permittivity $\varepsilon_1(\omega)$. The space separation
between the layers is $a$.
\\          
{\bf FIG.2.}& & 
The dielectric permittivity as a function of imaginary frequency for
$Al$ (solid line) and $Au$ (dashed line).
 \\
{\bf FIG.3.}& & 
The correction factor to the Casimir force due to finite conductivity of
the metal as a function of the surface separation.
The solid line 1 and 2 represents the computational results 
for $Al$ and $Au$ respectively in the configuration of two semi-spaces (a)
and for a sphere (lens) above a semi-space (b). The dashed lines 1 and 2
represent
the perturbation correction factor up to the 4th order for $Al$, and
$Au$ respectively.
\\
{\bf FIG.4.}& & 
The correction factor to the Casimir force due to finite conductivity of
the metal as a function of the surface separation
for $Al$ test bodies covered by thin layers of $Au$.
The dashed lines  
represent the results for a layer thickness $d=20\,$nm and the dotted lines 
for $d=30\,$nm. The case of the configuration of two semi-spaces is shown 
in (a)
and for a sphere (lens) above a semi-space is shown in (b). The solid lines 
represent the results for pure $Al$ and $Au$ test bodies respectively.
\\
{\bf FIG.5.}& & 
The absolute value of the van der Waals force as a function of surface 
separation is shown on a logarithmic scale.
The solid lines represent the results for $Al$ and 
the dashed lines represent the case of $Au$. The configuration of two 
semi-spaces is shown in (a)
and that for a sphere (lens) above a semi-space is shown in (b).
\end{tabular}
\begin{figure}[p]
\centerline{\epsffile{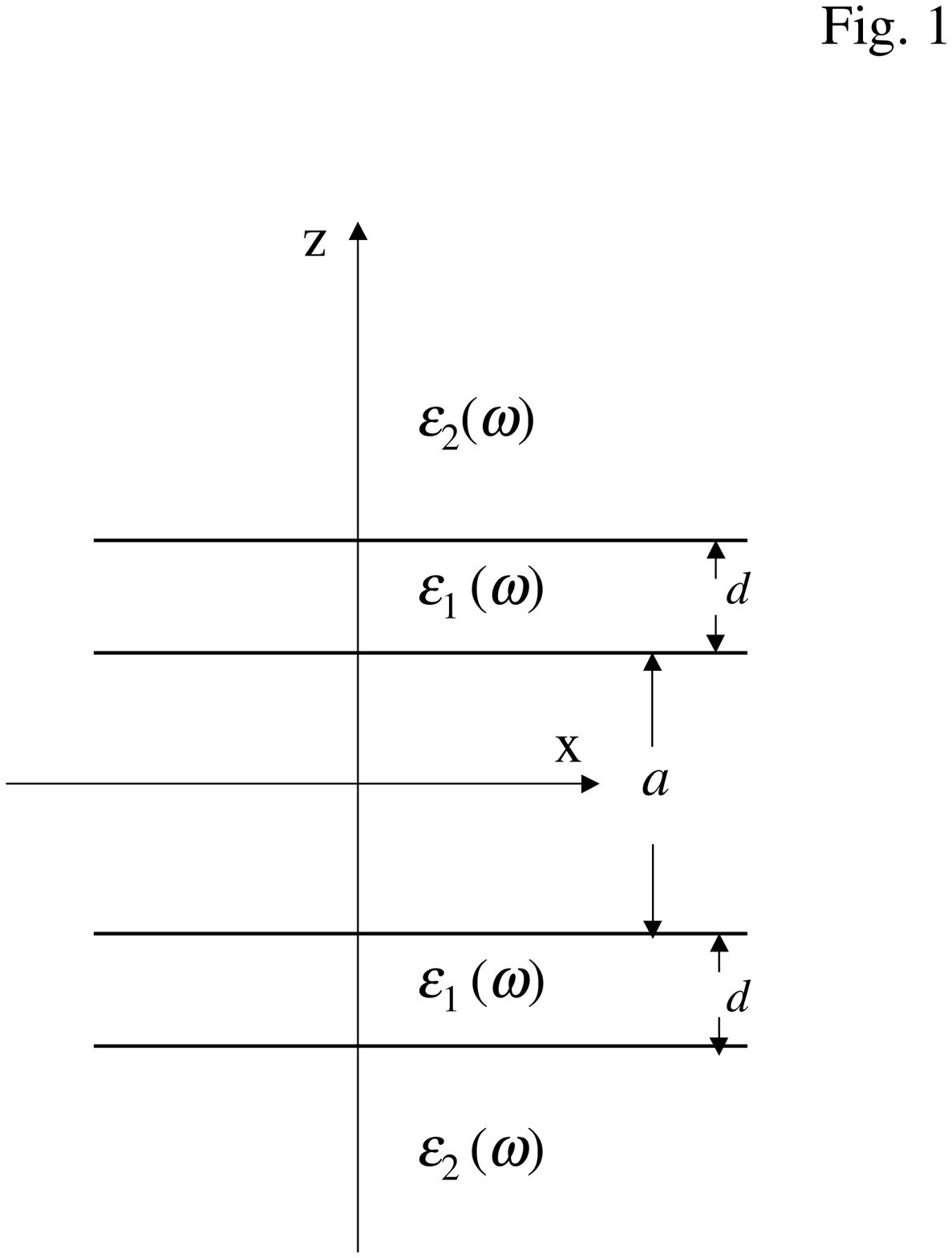} }
\end{figure}
\begin{figure}[p]
\centerline{\epsffile{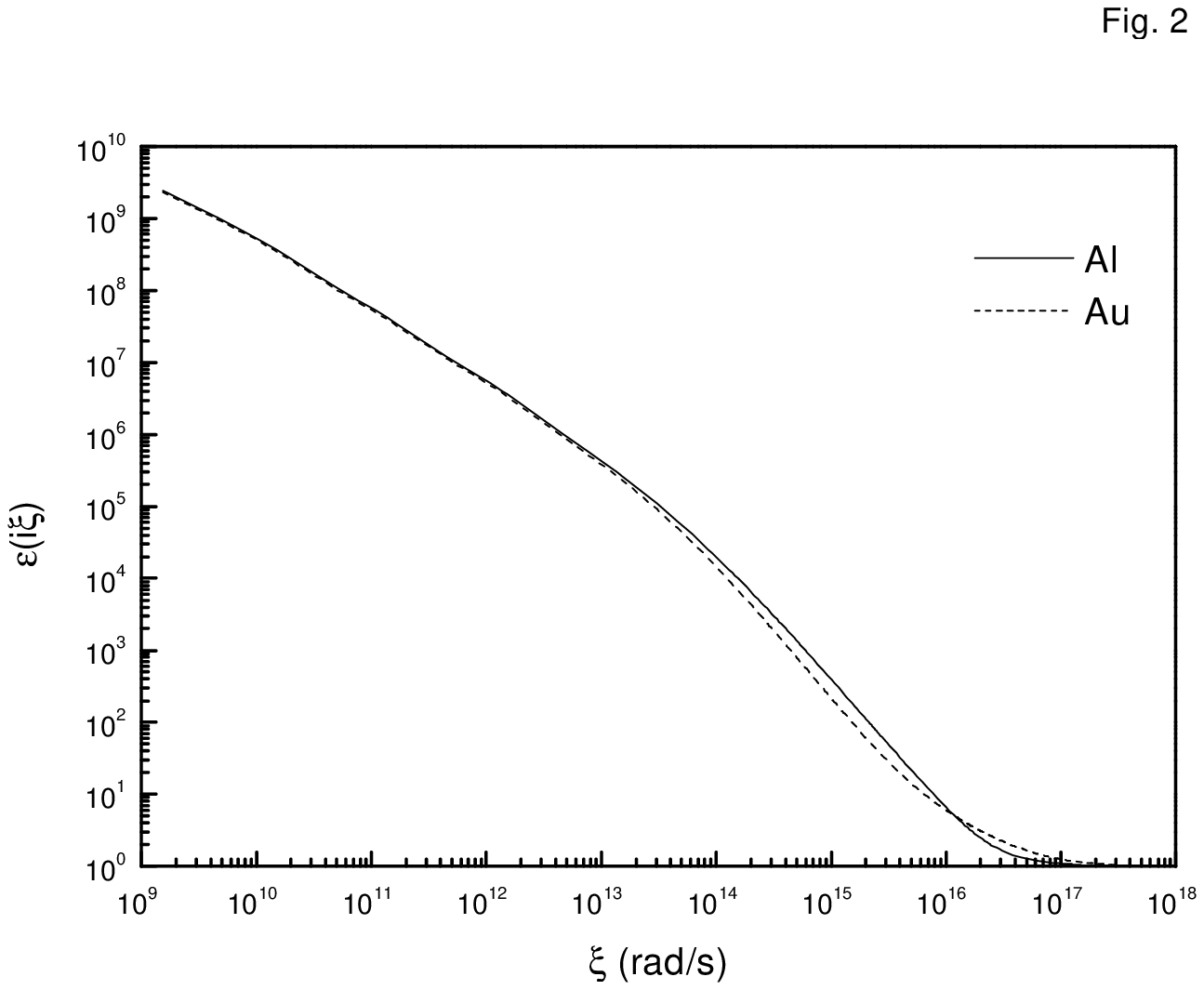} }
\end{figure}
\begin{figure}[p]
\centerline{\epsffile{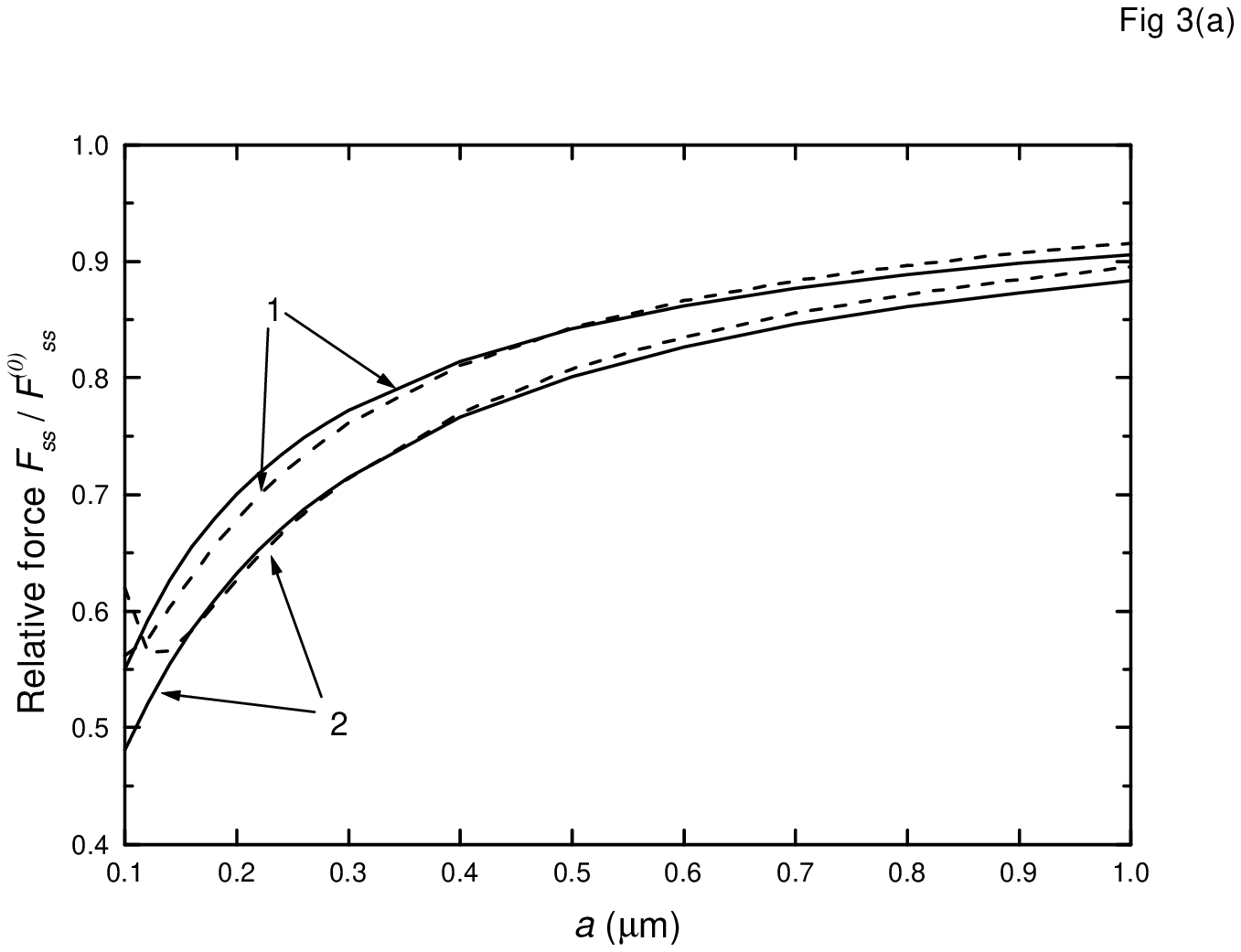} }
\end{figure}
\begin{figure}[p]
\centerline{\epsffile{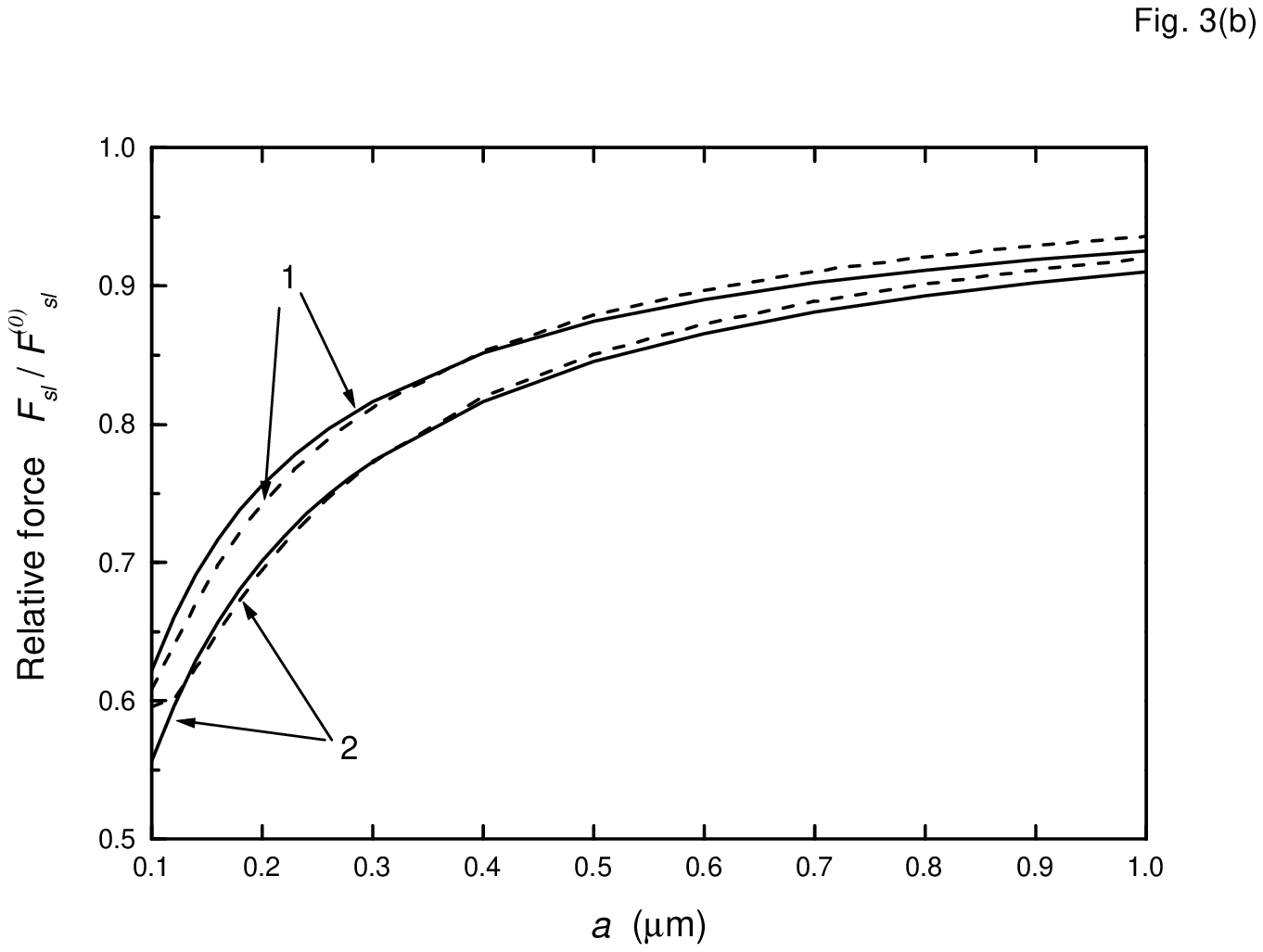} }
\end{figure}
\begin{figure}[p]
\centerline{\epsffile{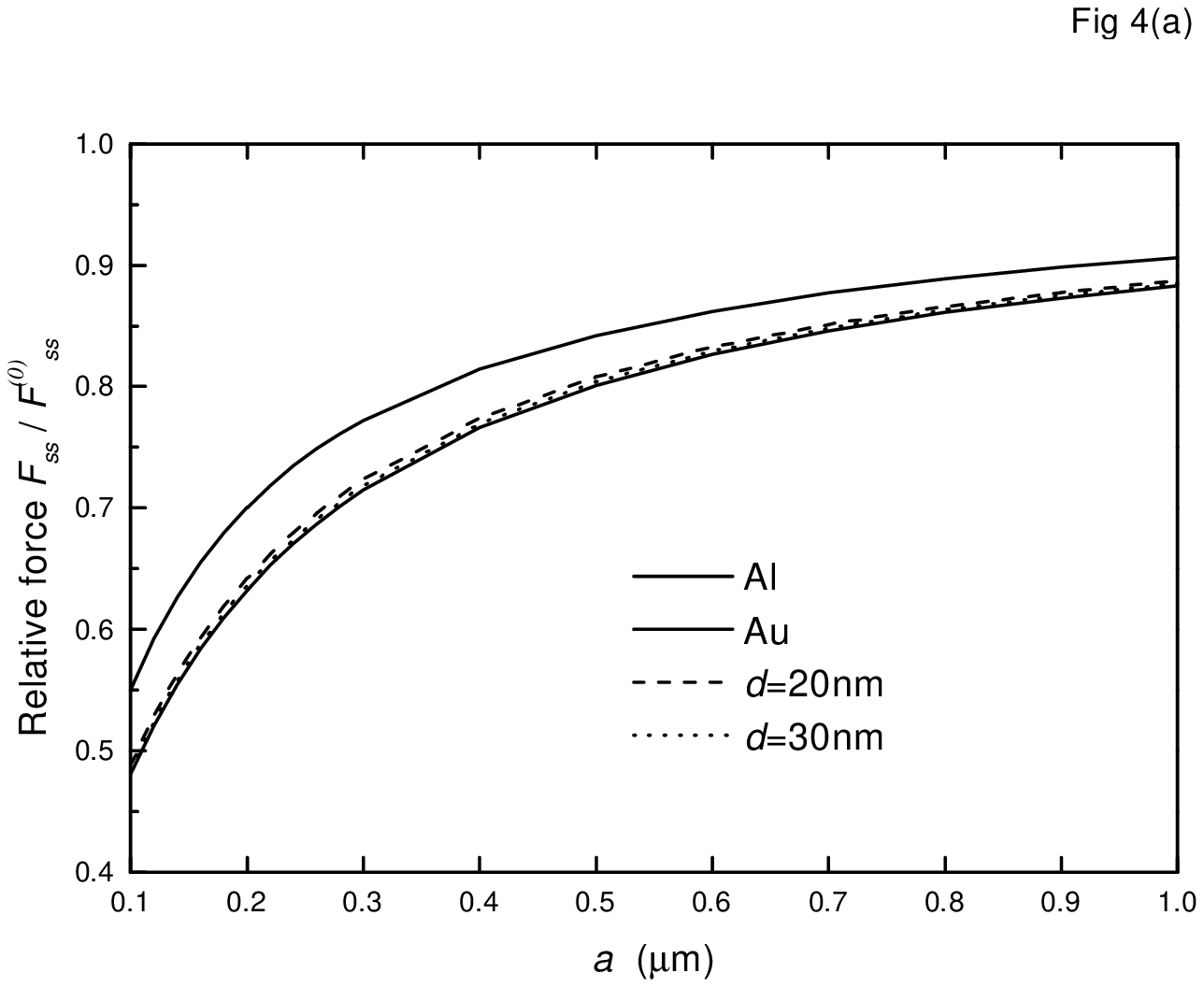} }
\end{figure}
\begin{figure}[p]
\centerline{\epsffile{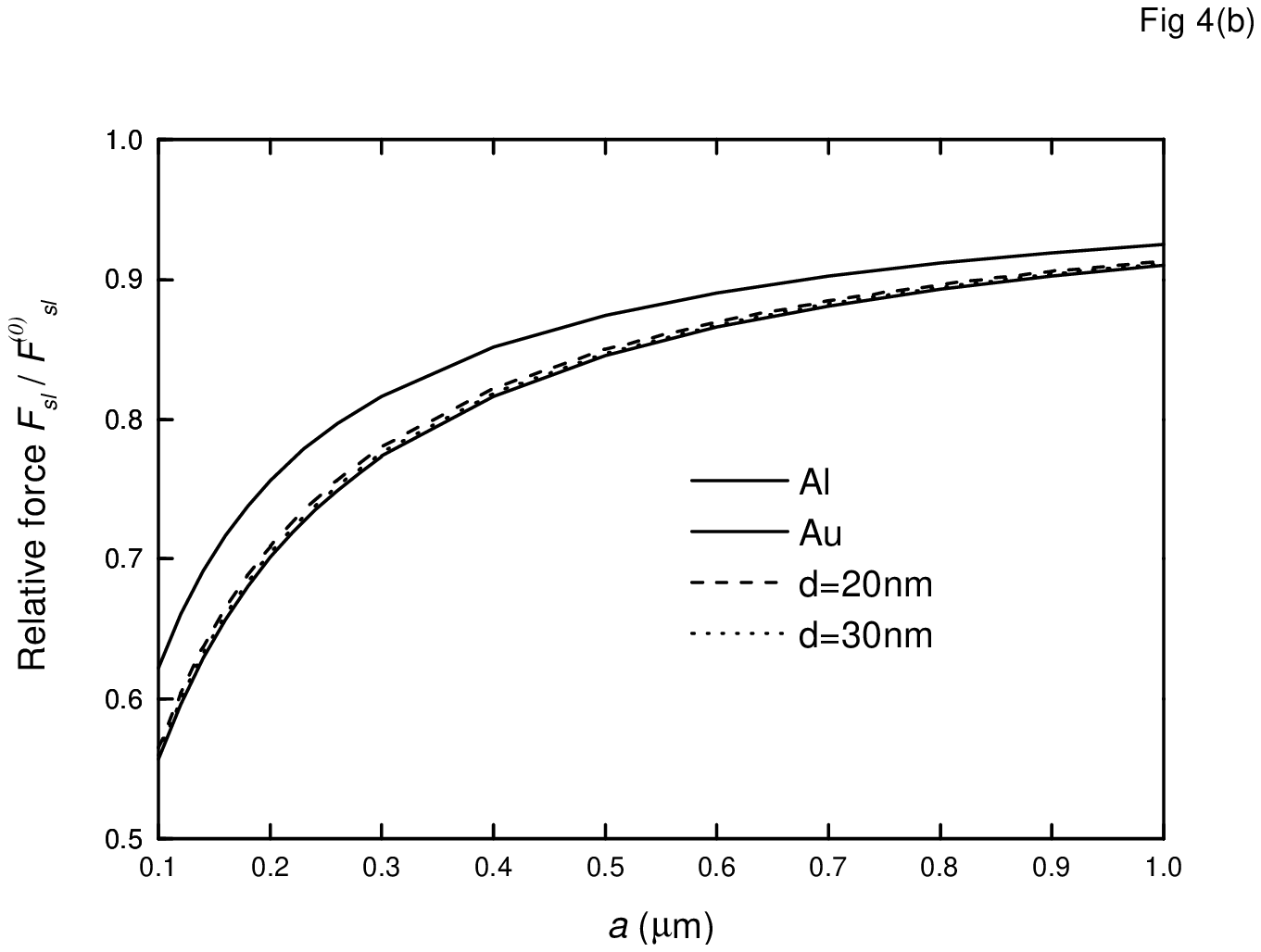} }
\end{figure}
\begin{figure}[p]
\centerline{\epsffile{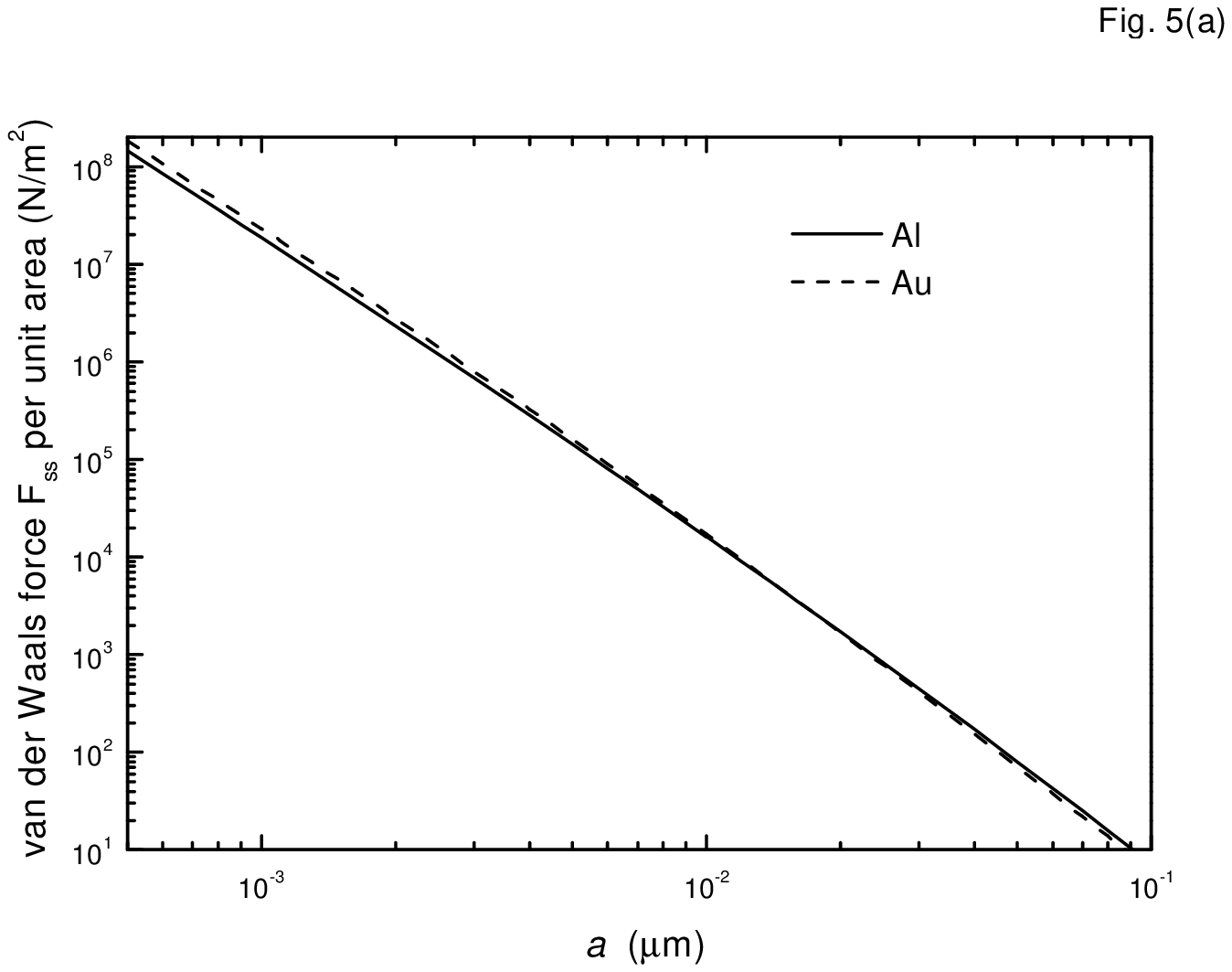} }
\end{figure}
\begin{figure}[p]
\centerline{\epsffile{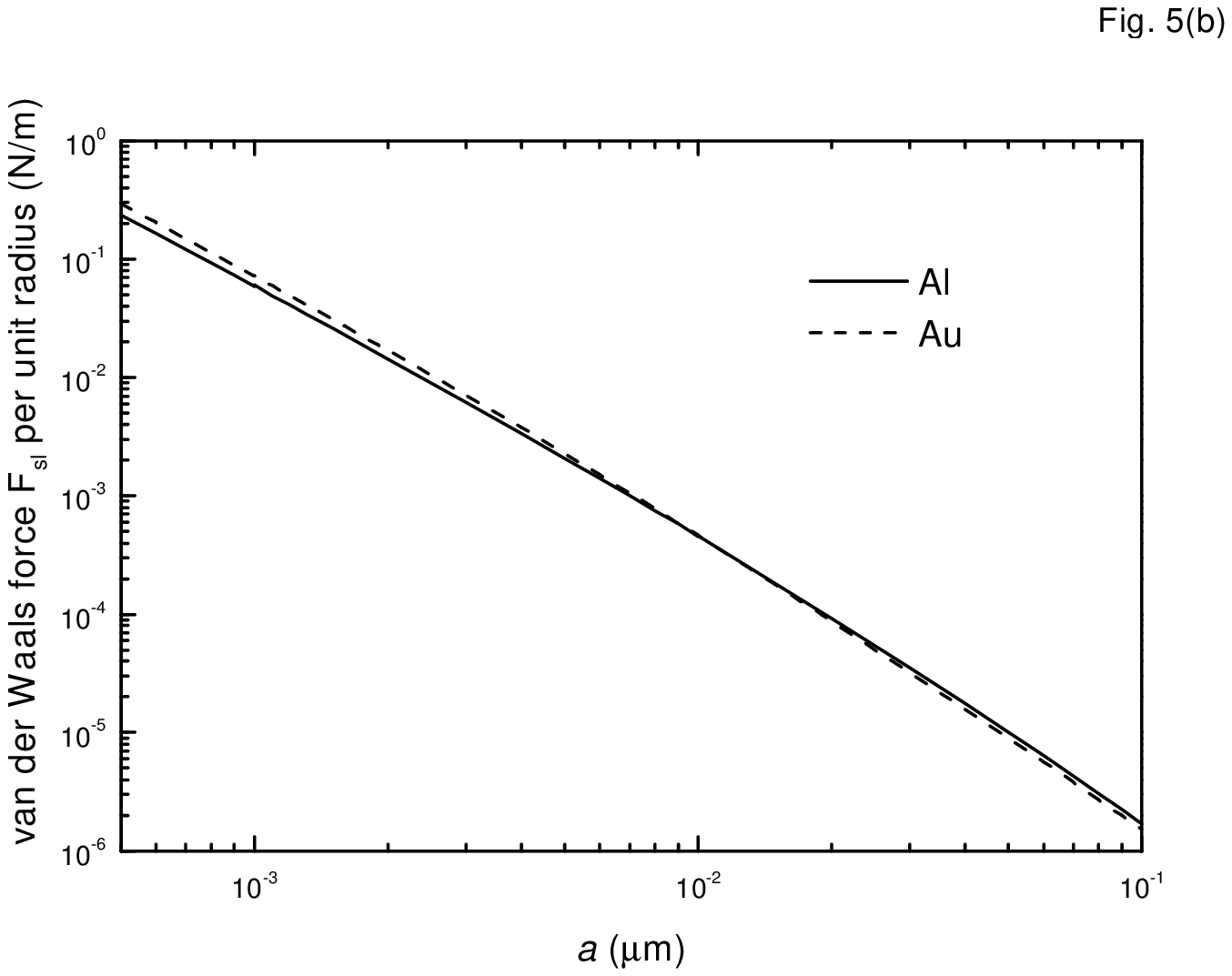} }
\end{figure}
\end{document}